\documentclass[onecolumn,showpacs,preprintnumbers,amsmath,amssymb,11pt]{revtex4-1}
\usepackage{amsmath}

\usepackage{graphicx}%
\usepackage[margin=1.8em,caption=false,subrefformat=parens]{subfig}  %
\usepackage{dcolumn}%
\usepackage{bm}%

\renewcommand{\d}{\mathrm{d}}
\newcommand{\unit}[1]{\ensuremath{\,\mathrm{#1}}}
\begin{document}

\title{Computing the demagnetizing tensor for finite difference
  micromagnetic simulations via numerical integration}

\author{Dmitri S. Chernyshenko}%
\author{Hans Fangohr}%
\affiliation{%
Engineering and the Environment\\
University of Southampton\\
SO17 1BJ Southampton\\
United Kingdom
}%

\date{\today}

\begin{abstract}
In the finite difference method which is commonly used in computational
micromagnetics, the demagnetizing field is usually computed as a convolution of the magnetization vector field with the demagnetizing tensor that describes the magnetostatic field of a cuboidal cell with constant
magnetization. An analytical expression for the demagnetizing tensor
is available, however at distances far from the cuboidal cell,
the numerical evaluation of the analytical expression can be very inaccurate.

Due to this large-distance inaccuracy numerical packages such as OOMMF compute the demagnetizing tensor
using the explicit formula at distances close to the originating cell, but at distances far from the originating cell a formula based on an asymptotic expansion has to be used. In this
work, we describe a method to calculate the demagnetizing field by
\emph{numerical evaluation of the multidimensional integral} in the
demagnetizing tensor terms using a sparse grid
integration scheme. This method improves the accuracy of computation at intermediate distances from the origin.

We compute and report the accuracy of (i) the numerical evaluation of
the exact tensor expression which is best for short distances, (ii)
the asymptotic expansion best suited for large distances, and (iii)
the new method based on numerical integration, which is superior to
methods (i) and (ii) for intermediate distances. For all three
methods, we show the measurements of accuracy and execution time as a
function of distance, for calculations using single precision
(4-byte) and double precision (8-byte) floating point
arithmetic. We make recommendations for the choice of scheme order and
integrating coefficients for the numerical integration method (iii).
\end{abstract}

\maketitle

\section{Introduction}
\label{sec:Introduction}

Micromagnetic simulation of ferromagnetic nanostructures is a
widespread tool to support research and device design in a variety of fields, including
magnetic data storage and sensing. The micromagnetic theory is based
on partial differential equations proposed in \cite{Brown1963a} combined
with an equation of motion that can be solved to determine the
time-development of the magnetization vector function. 

\subsection{Micromagnetics}

Numerical simulations in micromagnetics commonly solve the
Landau-Lifshitz-Gilbert and the associated partial differential
equations using a finite difference discretization of space (including
OOMMF, LLG Micromagnetics Simulator, Micromagus, Mumax, Micromagnum~\cite{oommf,llg,micromagus,mumax,micromagnum}).
In the finite difference method, space is discretized using a regular grid
with cuboidal cells, and the magnetization and other scalar and vector
fields involved in the computation are assumed to be constant within
each of these cubiodal cells. The magnetization vector field
$\mathbf{M}$ is the primary degree of freedom. As a function of
$\mathbf{M}$, which is represented by a constant within every cell,
various fields such as the exchange, anisotropy, Zeeman, and demagnetizing
fields are computed.
These are added together, and enter the equation of motion for $\mathbf{M}$ as the
effective field.

The most demanding part of the calculation is to determine the
demagnetizing field: to compute the demagnetizing field for one of
the discretization cells, an integral over the whole magnetic domain
has to be carried out, which translates into a (triple) sum (in a 3d
system) over all cells in the finite difference discretization. 

For a finite difference discretization of a three-dimensional sample
with $n_1$ discretization points in the x-direction, $n_2$ points in
the y-direction and $n_3$ points in the z-direction, there are $n =
n_1 n_2 n_3$ cuboidal cells in total. To compute the demagnetizing
field in each one of these cells, we need to consider the total
contribution of all $n$ cells. Thus, to work out the
demagnetizing field for all $n$ cells requires $\mathcal{O}(n^2)$
operations using a naive approach. For realistic mesh sizes
this is infeasible. Instead, usually the demagnetizing field
$\mathbf{H}$ is
expressed as a discrete convolution of the so-called demagnetizing tensor
$\mathbf{N}$ and the representation of the magnetization $\mathbf{M}$
\begin{equation}
\mathbf H_{i_1 i_2 i_3} = -\sum_{j_1 = 1}^{n_1}\sum_{j_2 = 1}^{n_2}\sum_{j_3 = 1}^{n_3} \mathbf N(\mathbf r_{i_1 i_2 i_3} - \mathbf r_{j_1 j_2 j_3}) \cdot \mathbf M_{j_1 j_2 j_3},
\label{eq:discreteconvolution}
\end{equation}
The triple indices $i_1 i_2 i_3$ are used to index cells in the
3d-discretization, i.e. $i_1$ counting cells in the x-direction, and
correspondingly $i_2$ and $i_3$ in y- and z- direction. The vector
$\mathbf{r}_{i_1 i_2 i_3}$ points to the centre of the cell $i_1 i_2 i_3$ and $\mathbf{M}_{i_1
  i_2 i_3}$ is the magnetization in that cell, and $\mathbf H_{i_1 i_2
  i_3}$ is the demagnetizing field in that cell. 

The discrete
convolution (\ref{eq:discreteconvolution}) is typically carried out as a product in Fourier space as
the regular spacing of the finite difference cells allows straightforward 
use of the Fast Fourier Transform and its inverse. The
demagnetizing tensor~$\mathbf N$ needs to be computed once for
given geometry and discretization, normally at the setup stage of the
simulation. In this work we propose a new procedure for the accurate computation
of entries in the demagnetizing tensor.

The tensor $\mathbf N(\mathbf r)$ describes the
energy of the demagnetizing interaction between two uniformly
magnetized cuboids~$\sigma_1$ and~$\sigma_2$ of volume~$|\sigma|$ separated by the translation vector $\mathbf r$. It is a symmetric tensor
of rank 2, which we write in the dimensionless form following the convention from~\cite{Newell1993}
\begin{align}
\mathbf N(\mathbf r) &= - \frac{1}{4 \pi |\sigma|} \int_{\sigma_1} \d{\mathbf r_1}
\int_{\sigma_2(\mathbf r)} \nabla_{{\mathbf r}_1} \nabla_{\mathbf r_2}
\frac{1}{|\mathbf r_1 - \mathbf r_2|} \, \d\mathbf r_2
\label{eq-n} \\
E_{\sigma_1 \leftrightarrow \sigma_2}&= \mu_0 |\sigma| \, \mathbf M_1 \cdot \mathbf N(\mathbf r) \cdot \mathbf M_2
\label{eq-demag-energy}
\end{align}

The computation of the demagnetizing field using the formula~\eqref{eq:discreteconvolution} follows the commonly used energy-based approach to the discretization of the Landau-Lifshitz-Gilbert equation~\cite{Miltat2007}. In this approach (used in OOMMF~\cite{oommf} and in our work), the components of the effective field in each cuboidal cell are obtained via a minimization procedure applied to the discretized total energy~$E_\mathrm{total} = \sum_{\sigma_1, \sigma_2} E_{\sigma_1 \leftrightarrow \sigma_2}$. In contrast, in the field-based approach the discretized fields are obtained from the corresponding continuous fields by e.g. sampling at the cell center~\cite{Labrune1995,Albuquerque2002,Toussaint2002,Miltat2007}, and the demagnetizing tensor~\eqref{eq-n} is not used.

\subsection{Numerical accuracy of $\mathbf N(\mathbf r)$ evaluation}

The components of the demagnetizing tensor can be computed via an analytical formula~\cite{Schabes1987,Newell1993,maicus1998magnetostatic,fukushima1998volume}. However, when $\mathbf r =
\mathbf{r}_a - \mathbf{r}_b$ is large compared to the size of the mesh
cells (i.e. when the interacting mesh cells at $\mathbf{r}_a$ and
$\mathbf{r}_b$ are far apart on the grid), evaluation of this expression on a computer can result in a loss of significant digits, to a point where the computed answer contains no significant digits at all~\cite{Donahue2007}. 

The loss of accuracy is caused by catastrophic cancellation: the terms of the analytical expression correspond to indefinite integrals with $r^3$ order of growth, while the demagnetizing tensor itself (the definite integral) is of the order $1/r^3$. On modern CPUs, double-precision floating numbers contain approximately 15 significant digits, therefore the relative error in the computation of the demagnetizing tensor using the analytical formula is of the order $10^{-15} r^6$. One can therefore expect that for cell separations greater than $10^{15/6} \approx 300$ the analytical computation will contain no significant digits at all (i.e., the relative error will be greater than 1).

Indeed, the above estimate is confirmed if the result of the computation using the analytical formula is compared to the exact (up
to machine precision) value of the demagnetizing tensor. 
The exact value is computed using specialized high-precision libraries
(see Sec. \ref{sec:error-estimation} and \ref{sec:other-high-accuracy}).
As seen in Fig.~\ref{fig-final-comparison}, the relative error $\eta$ of the analytical computation grows as $r^6$ and crosses the $\eta = 1$ threshold at a separation of about 300 cells.      

The micromagnetic simulation package OOMMF counteracts this inaccuracy problem by
utilizing an asymptotic expansion of the demagnetizing tensor \cite{Donahue2007} in terms of powers of $1/r$ up to 6th order.
In this paper we investigate an alternative approach to deal with the catastrophic cancellation problem, which is to compute the
integral (\ref{eq-n}) directly using numerical integration. 

\section{\label{sec:Motivation}Method}
As outlined in section~\ref{sec:Introduction}, the 6-d integral described
in~\eqref{eq-n} can be computed using
\begin{itemize}
  \item an analytical formula~\cite{Newell1993},
  \item an asymptotic expansion~\cite{Donahue2007},
  \item numerical integration.
\end{itemize}

For computing integrals in one dimension, multiple highly-accurate
methods are available.
However, the computation of multidimensional integrals is hindered by
the so-called curse of dimensionality, where the number of integration
points increases exponentially with the increase in dimension. Since
the demagnetizing tensor~$\mathbf N(\mathbf r)$ has to be computed for
all possible grid offsets~$\mathbf r$, the integration method for this
six-dimensional integral needs to be both accurate and fast. 

The tradeoff between accuracy and computational effort can be achieved
using the sparse grid family of methods, also known as Smolyak
quadrature~\cite{Smolyak1963} (for a brief review of other
alternatives see~\cite{Petras2003}).

The sparse grid method is used to extend one-dimensional integration rules to integration formulas in multiple dimensions. Below we summarize the key ideas of the method~\cite{Petras2003}. Starting with a one-dimension family of integration formulas $I_k$ for computing the 1d integral
\begin{equation}
\int_0^1 f(x) dx \approx I_k[f] = \sum_{i=1}^{n_k} a_{ki} f(x_{ki})
\label{eq-1drule} 
\end{equation}
we can write the formal identity
\begin{align*}
\int_0^1 f(x) dx &= I_0[f] + (I_1[f] - I_0[f]) + \ldots \\ 
&= \Delta_0[f] + \Delta_1[f] + \ldots
\end{align*}
where $\Delta_k = I_k - I_{k-1}$ and $\Delta_0 = I_0$.

An example for a family of integration formulas $I_k$ are Gaussian integration formulae with $k+1$ points.

Now, we apply the above formal identity $d$ times to obtain a $d$-dimensional integration rule
\begin{equation}
\int_{[0,1]^d} f(\mathbf x) d\mathbf x = \Big( \prod_{j=1}^d (\Delta_0^{(j)} + \Delta_1^{(j)} + \ldots) \Big)[f]
\label{eq-ndrule} 
\end{equation}
Here the $\Delta_k^{(j)}$ operator applies the
approximation~\eqref{eq-1drule} to the $j$-th argument of the function
$f$, and the $\prod$ symbol represents the operator product, i.e. repeated application of $\sum_k \Delta_k^{(j)}$ in all $d$ dimensions.
 
 The formal expansion~\eqref{eq-ndrule} is infinite; to obtain a practical integration formula we need to truncate it. Smolyak quadrature achieves this by expanding the product~\eqref{eq-ndrule} and grouping together terms $\Delta_{i_1}^{(1)} \cdots \Delta_{i_d}^{(d)}$ with the same term order 
 $k = i_1 + \ldots + i_d$:
\begin{align}
\int_{[0,1]^d} f(\mathbf x) d\mathbf x 
	&= Q_0[f] + Q_1[f] + \ldots \\
	&\approx Q_0[f] + Q_1[f] + \ldots + Q_k[f] \label{eq-smolyak}
\end{align}
where
\begin{equation}
Q_k[f] = \sum_{i_1 + \ldots + i_d = k} \Delta_{i_1}^{(1)} \Delta_{i_2}^{(2)} \cdots \Delta_{i_d}^{(d)}[f]
\end{equation}
Compared to the evaluation of the $d$-dimensional integral using a naive $d$-fold product (which required $n_k^d$ integration points for rule $I_k$), the Smolyak formula~\eqref{eq-smolyak} greatly reduces the number of integration points required to achieve a certain level of accuracy, as long as the integrand is reasonably smooth.

Different one-dimensional families~\eqref{eq-1drule} will result in different
multidimensional formulas~\eqref{eq-smolyak}; in our testing for the
demagnetizing integrand~\eqref{eq-n} we obtained the best results when using
the ``delayed Kronrod-Patterson sequence'' developed by Petras~\cite{Petras2003}. The delayed sequence is based on the Kronrod-Patterson family of 1d integration formulas $I_k^{\mathrm{KP}}$~\cite{Patterson1968}, however some of the formulas are repeated to lower the rank of approximation (and the required number of integration points), determined by the ``delay sequence`` $k_i$:
\begin{equation}
I_i^{\mathrm{delayed}} = I_{k_i}^{\mathrm{KP}}
\end{equation}
For the maximum delay sequence~\cite{Petras2003} the formulas are repeated so that the 1-d rule $I_i^{\mathrm{delayed}}$ is accurate
for polynomials up to rank~$i$:    
\begin{equation}
k_i^{\mathrm{full}} = 0,1,1,2,2,2,3,3,3,3,3,3,4,\ldots
\end{equation}

 \subsection{Integrand}

\subsubsection{6d method}
\label{sec:6d-method}
A straightforward way to compute the demagnetizing tensor numerically
would be to apply the sparse grid formulas to the 6-d
integral~\eqref{eq-n}. 

\subsubsection{4d method}
\label{sec:4d-method}
Due to the high dimensionality of the 6d integral,
the number of required integration points will be quite high. To
reduce the dimensionality, we can transform the 6-d volume
integral~$\eqref{eq-n}$ to a 4-d surface integral using
a variant of Gauss's theorem. The procedure is described in~\cite{Newell1993} and results in the following formulas for the components of the
demagnetizing tensor where we have used the notation of~\cite{Newell1993}
\begin{equation*}
N_{xx}(X,Y,Z) = \frac{1}{4\pi \Delta x \, \Delta y \, \Delta z} \big[ 2F(X,Y,Z) -
F(X+\Delta x, Y,Z) - F(X-\Delta x,Y,Z)\big]
\end{equation*}
\begin{equation*}
N_{xy}(X,Y,Z) = \frac{1}{4\pi \Delta x \, \Delta y \, \Delta z} \big[ G(X,Y,Z) 
- G(X-\Delta x,Y,Z) 
- G(X,Y+\Delta y,Z) 
 +G(X-\Delta x,Y+\Delta y,Z) \big]
\end{equation*}
with
\begin{equation*} 
F(X,Y,Z) =
 \int\limits_0^{\Delta z}\int\limits_0^{\Delta y} \int\limits_0^{\Delta z} \int\limits_0^{\Delta y}
\frac{dz\, dy \, dz' \, dy'}{
\sqrt{X^2 + (y+Y-y')^2 + (z+Z-z')^2}
}
\end{equation*}
and
$$
G(X,Y,Z) = 
\int\limits_{Y-\Delta y}^Y \int\limits_{Z-\Delta z}^{Z} \int\limits_z^{z+\Delta z} \int\limits_X^{X+\Delta x}
\frac{dy \, dz \, dz' \, dx'}{
\sqrt{x'^2 + y^2 + z'^2}
}
$$
where $(X, Y, Z) = \mathbf{r}$ is the vector between the two
interacting cells, and $\Delta x$, $\Delta y$, $\Delta z$ are the edge lengths of the cuboidal cells.

To simplify the application of numerical integration formulas, we transform the above expressions so that integration is performed over the unit cube $[0,1]^4$.
\begin{equation}
N_{xx}(X,Y,Z) = 
	\int\limits_0^1\int\limits_0^1 \int\limits_0^1\int\limits_0^1   
	\big[ 2\mathcal{F}(X,Y,Z)  
	- \mathcal{F}(X+\Delta x, Y,Z) - \mathcal{F}(X-\Delta x,Y,Z)\big]
\,	dz\, dy \, dz' \, dy' 
\label{eq-Nxx-4d}
\end{equation}
\begin{multline}
N_{xy}(X,Y,Z) = 
	\int\limits_0^1\int\limits_0^1 \int\limits_0^1\int\limits_0^1 dy \, dz \, dz' \, dx' 
	\times \\
 \times \big[ \mathcal{G}(X,Y,Z)   
 - \mathcal{G}(X-\Delta x,Y,Z) 
 - \mathcal{G}(X,Y+\Delta y,Z) 
 +\mathcal{G}(X-\Delta x,Y+\Delta y,Z) \big]
\label{eq-Nxy-4d}
\end{multline}
with
\begin{equation*}
\mathcal{F}(X, Y, Z) = \frac{\Delta y \, \Delta z}{4\pi \, \Delta x}  
 \frac{1}{\sqrt{X^2 + (\Delta y \, y+Y-\Delta y \, y')^2 + (\Delta z \, z+Z-\Delta z \, z')^2}}
\end{equation*}
\begin{equation*}
\mathcal{G}(X, Y, Z) = \frac{\Delta z}{4\pi}   
\frac{1}{\sqrt{(X + \Delta x \, x')^2  
+ (Y +\Delta y (y-1))^2 + (Z + \Delta z (z + z' - 1))^2}}
\end{equation*}
The remaining components of the tensor can be obtained by variable substitution
(for example, $N_{yz}(X,Y,Z) = N_{xy}(Y,Z,X)$).

In this approach, we are essentially computing two dimensions of the 6-d integral analytically, and the remaining four numerically. As we shall see, the two analytical steps introduce a small amount of cancellation, but the required number of integration points is significantly reduced.
  
\subsection{Error estimation}
\label{sec:error-estimation}

To determine the accuracy of the computed demagnetizing tensor $\mathbf N$, we
compute its exact (to machine precision) value $\mathbf N_\text{exact}$ via the analytical formula using the GNU MPFR high-precision floating point library~\cite{Fousse2007}, and evaluate the relative error $\eta$:
\begin{equation}
\label{eq:eta}
\eta = \frac{||\mathbf N - \mathbf N_\text{exact}||}{||\mathbf
N_\text{exact}||}
\end{equation}
where the matrix norm is defined as $||\mathbf N|| = \sqrt{\sum_{i,j} N_{ij}^2}$.

\section{Results}

\begin{figure} \centering
\includegraphics[width=10cm]{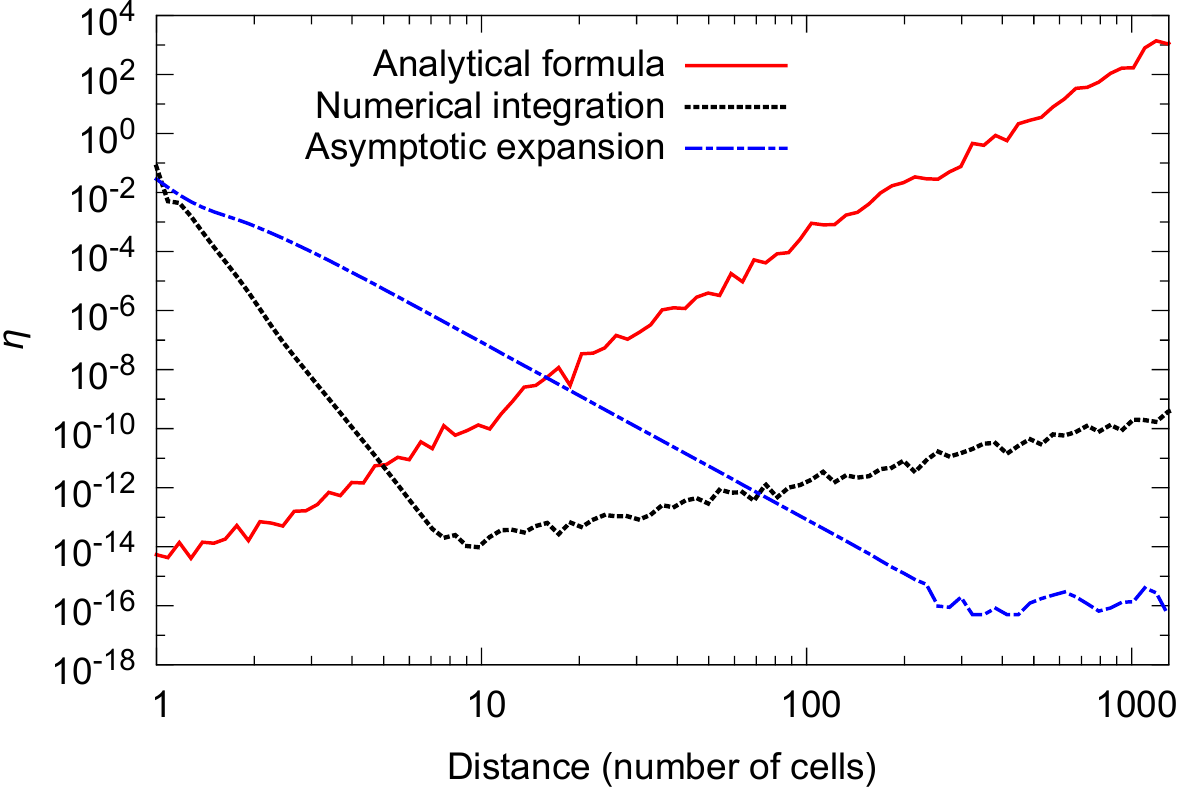}%
\caption{The relative error $\eta$ in the computation of the
  demagnetizing tensor as a function of the distance between the
  interacting cuboids, $1\!\times\! 1\!\times\! 1$ cell size, Kronrod-Patterson
  sparse grid integration with full delay, order $k=7$~\cite{Petras2003},
  double precision arithmetic (precision $\sim 10^{-16}$).
}%
\label{fig-final-comparison} 
\end{figure}

\subsection{Overview}
\label{sec:overview}

Figure~\ref{fig-final-comparison} shows the comparison of the accuracy
of the analytical formula, the asymptotic expansion, and the numerical
sparse grid integration methods as a function of the distance between cells using (8 byte) double floating point.
For the numerical sparse grid integration, the 4d method (Sec. \ref{sec:4d-method}) has been used.

While the aspect ratio of
the cuboidal cell with edge lengths $\Delta x, \Delta y$ and $\Delta
z$ affects the results somewhat, they are independent of the absolute
size of the cuboidal cell. We have chosen $\Delta x, \Delta y$ and
$\Delta z = (1, 1, 1)$ so that the distance $|\mathbf{r}|$ between
interacting cuboids is expressed in the number of cells between the
interacting cuboids. For example, for a micromagnetic simulation with a
$5\unit{nm} \times 5\unit{nm} \times 5\unit{nm}$ cell size, the distance of 10
on the plot in Figure~\ref{fig-final-comparison} corresponds to a $50\unit{nm}$
distance on the mesh.

We start by discussing the analytical formula shown as a
red solid line in Fig.~\ref{fig-final-comparison}. Its relative error $\eta$ for very short distances is
$10^{-14}$. We cannot expect an error below $10^{-16}$ as this is
the precision of the double floating point numbers
used. As the distance between interacting cells increases, the analytical
formula becomes less accurate. At a separation of 100 cells, it is about
$10^{-4}$, meaning that only the first 4 digits are correct. In fact, beyond a
distance of about 300 cells the relative error becomes greater than 1, indicating that no digits of the double float
can be expected to be correct and that not considering the demagnetizing
tensor beyond that point would be more accurate than computing it
analytically. 

The asymptotic expansion (blue dash-dotted line in Fig.~\ref{fig-final-comparison}) starts with a large
relative error $\eta \sim 10^{-2}$ for short distances, which
decreases as the distance increases. At about 200 cells distance, the
relative error is $\sim 10^{-15}$ and remains of that magnitude for
larger distances. The smooth reduction of the error with distance
reflects the way the high-order moments of the cuboid interaction
decay with increasing distance, thus making the asymptotic
approximation increasingly more accurate. The asymptotic expansion is
more accurate than the exact analytical expression for distances greater
than about 11 cells.

The precise number of cells for which this crossover occurs depends on
the aspect ratio of the discretization cell as well as the direction
of the separation vector between the two interacting cuboids. In
practical implementations of finite difference micromagnetic codes a
crossover point needs to be identified. In OOMMF, by default the 
asympotic formula is used for distances above 32 cells, which for this
example corresponds to an error of $10^{-6}$.

The numerical sparse grid integration error (black dotted line in Fig.~\ref{fig-final-comparison}) also starts around
$10^{-2}$ for short distances and decays to $~10^{-14}$ for a cell
distance of about 7. For very short distances, the integrand
\eqref{eq-n} varies quickly (it diverges for $|\mathbf{r}|
\rightarrow 0$) and
numerical integration is inaccurate. For cell distances between 7 and 70,
numerical integration is more accurate than the analytical expression, and more accurate than the asymptotic expansion. 
Beyond radius 70, the asymptotic expansion is more accurate than numerical integration. 
The slight inrease in the relative error of the numerical formula with
increasing distance is caused by the cancellation introduced by Gauss's theorem
(see~\ref{sec:4d-method}).

In summary, the numerical evaluation of the analytical formula is most
accurate for short distances, and the asymptotic expansion is most accurate for
long distances. The new sparse grid integration method introduced here is most
accurate for intermediate distances.

\subsection{Sparse grid integration parameters and execution performance}

\begin{figure}[htb] \centering
\includegraphics[width=10cm]{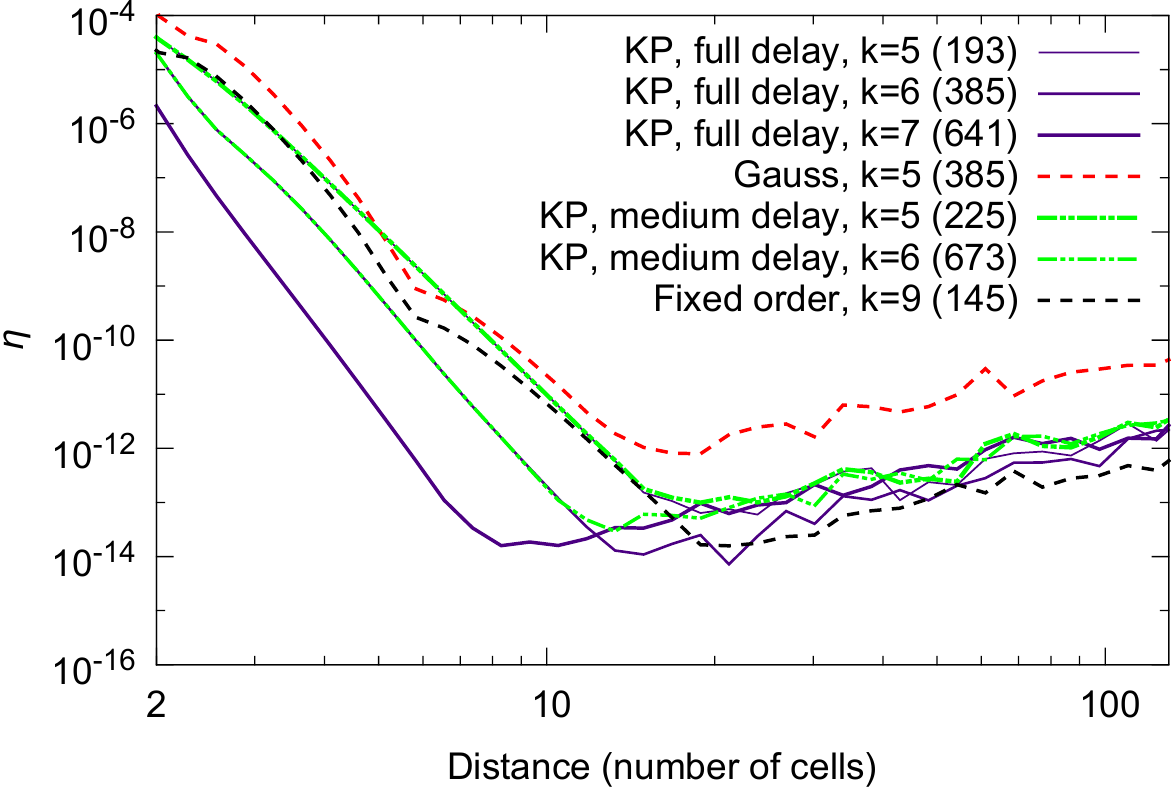}%
\caption{Comparison of numerical integration schemes for computing the
4d integral~\eqref{eq-Nxx-4d}-\eqref{eq-Nxy-4d}; KP: sparse
grid integration based on the delayed Kronrod-Patterson 1d rule~\cite{Petras2003};
Gauss: sparse grid integration based on the delayed Gauss 1d rule; fixed order:
9th order rule with 145 integration points~\cite{ecf,Sorevik85}; $k$: the
rule's order of approximation; ($\cdot$): the number of integration points.
}%
\label{fig-number-of-points} 
\end{figure}

In the sparse grid method, the order of approximation is a parameter
that can be adjusted to reach the desired level of accuracy or
performance. For lower orders of approximation, one can also use fixed-order
integration formulas that usually provide the same accuracy with fewer
integration points. An extensive library of such formulas is available~\cite{ecf,Cools2003,Cools1999,Cools1993}.

Since the computation of the demagnetizing tensor is a one-off cost, higher order,
more accurate integration formulas would be preferable. In
Figure~\ref{fig-number-of-points} we show the results for a number of sparse
grid formulas as well as for a 145-point fixed-order formula
from~\cite{ecf,Sorevik85}.

The most accurate method considered here is the 641-point 7th order sparse grid
method based on the 1d Kronron-Patterson sequence with full
delay~\cite{Petras2003}.
The delayed Kronrod-Patterson methods with orders 6 and~5 have fewer
points and decreased accuracy. Formulas based on Gauss or Kronrod-Patterson
sequences with medium delay require extra integration points to
achieve the same level of accuracy and are thus suboptimal compared to the
Kronrod-Patterson family with full delay.

The 145-point fixed order formula is more accurate than the 193-point 5th order
Kronrod-Patterson formula, but slightly less accurate than the 385-point 6th
order formula.

Based on these results, we recommend using either the 145-point fixed order
formula or the Kronrod-Patterson fully delayed formula with order 6 or 7,
depending on required accuracy and performance.

\subsection{Comparison of 4d and 6d integration}

\begin{figure} \centering
\includegraphics[width=10cm]{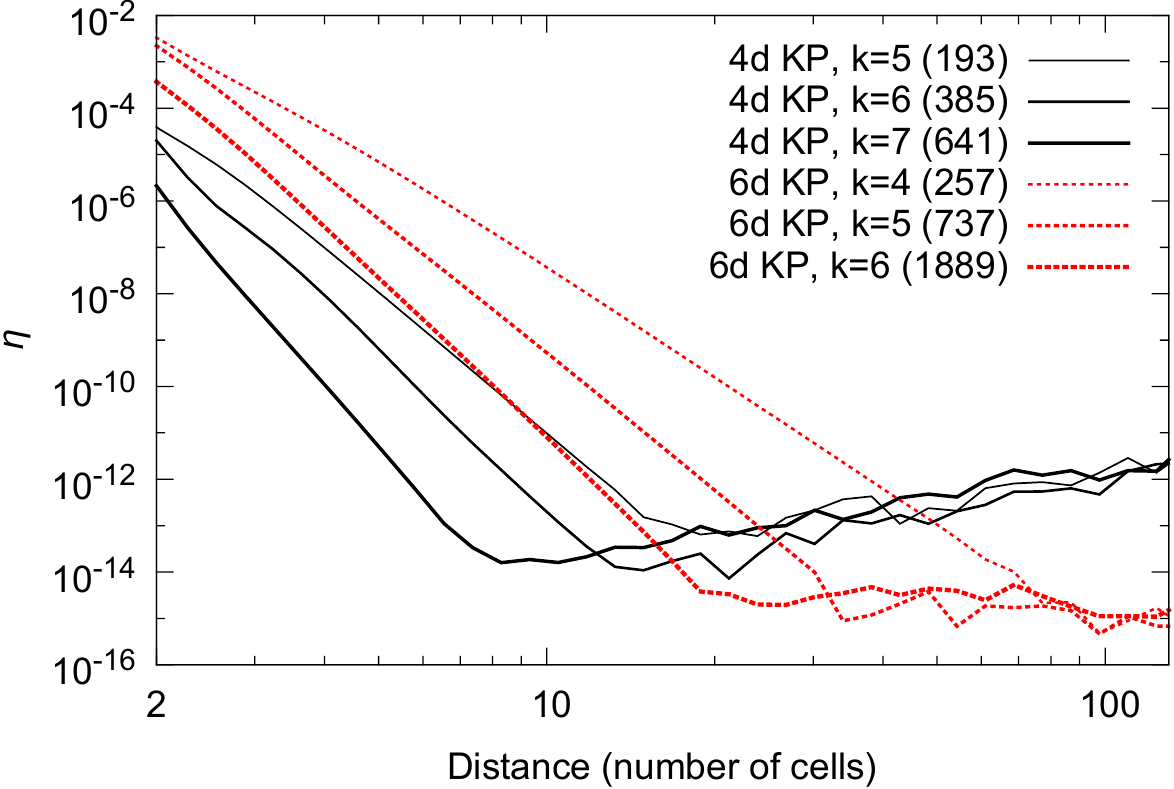}%
\caption{Comparison of numerical integration of the 4d integral~\eqref{eq-Nxx-4d}-\eqref{eq-Nxy-4d} versus the
6d integral~\eqref{eq-n}, Kronrod-Patterson 1d rule with full delay, $1\!\times\! 1\!\times\! 1$ cells.}%
\label{fig-4d-vs-6d} 
\end{figure}

As an alternative to computing the 4d integral (method~\ref{sec:4d-method}),
we can also compute all of the 6 integrations numerically
(method~\ref{sec:6d-method}). The results for the 6d
integration method are shown in Figure~\ref{fig-4d-vs-6d}. All red dashed curves show results
for the 6d method, and all black solid lines show results for the 4d
method. 

Both for the three 4d data sets and the three 6d data sets, we can see
that the decrease in the error with increasing distance is faster for higher
order methods. The 4d lines show minimal error
between 8 and 11 cells distance, and for larger distances the error increases a
little --- this reflects the numerical cancellation from subtracting large terms that increase with distance and originate from the analytical
integration that has been carried out over 2 dimensions. On the
contrary, the 6d data sets --- where the whole 6d integral has been
computed numerically --- does not suffer from this and the error remains
small ($<10^{-14}$) for larger distances. However, the number of
evaluation points is higher compared to the 4d method.

As seen in Figure \ref{fig-number-of-points}, even the lowest order
quadrature formula gives a sufficient accuracy gain in the
intermediate range to be better than the analytic and the asymptotic
expression. However, in time-dependent micromagnetic simulations the
computation of the demagnetizing tensor is a one-time setup cost, and
the setup cost is only slightly influenced by the order of the
integration formula --- one might as well choose the higher order, more accurate
formula.

Comparing the 4d and 6d method, we suggest the 4d method as it is
faster to compute. The 4d method is less accurate than
the 6d for the largest distances but this is of little practical concern
--- for those distances, the asymptotic expansion can be used instead.

\subsection{Performance}

In table~\ref{table-performance}, we show performance measurements for computing the 
entries in the demagnetizing tensor. For the fixed order (k=9) 145-point integration rule, the computational cost is comparable to the analytical formula, while the KP full delay (k=7) 641-point rule is approximately 5 times slower. The total cost of evaluating the demagnetizing tensor on a $400 \times 40 \times 1$ mesh was $95\unit{ms}$ (Table~\ref{table-performance2}) for the 145-point rule and $351\unit{ms}$ for the 641-point rule. Since a typical dynamical micromagnetic simulation usually requires 10,000 or more time steps, the one-time cost of setting up the demagnetizing tensor is minor even for the 641-point rule. For the sample system studied here, evaluating the effective field and dm/dt once takes $4\unit{ms}$. We also note that as the mesh grows larger, more of the entries can be computed using the extremely fast asymptotic formula, resulting in sublinear scaling of the total cost with mesh size $n$.       

\begin{table}[htb]
\begin{tabular}{lc}
Method & Time per cell, ns \\
\hline
Analytical                &  8.5 \\
Integration, 145 points  &     9.8 \\ 
Integration, 641 points &  43.2 \\
Asymptotic                 & 0.2 \\

\end{tabular}%
\caption{Per-cell cost of computing the entries of the demagnetizing tensor, workstation: dual CPU Intel E5506 2.13 GHz (8 threads), compiler: GCC 4.7.2.
 \label{table-performance}}%
\end{table}

\begin{table}[htb]
\begin{tabular}{lc}
Method & Total time, ms \\
\hline
Combined, 145 points       &    95 \\
Combined, 641 points     &  351 \\
\hline
LLG dm/dt evaluation    &       4

\end{tabular}%
\caption{Total cost of computing the demagnetizing tensor for a $2000 \times 200 \times 20\unit{nm}^3$ mesh, $5 \times 5 \times 20\unit{nm}$ cells.   
 \label{table-performance2}}%
\end{table}
      
\subsection{Single point floating precision}

\begin{figure}[htb] \centering
\includegraphics[width=10cm]{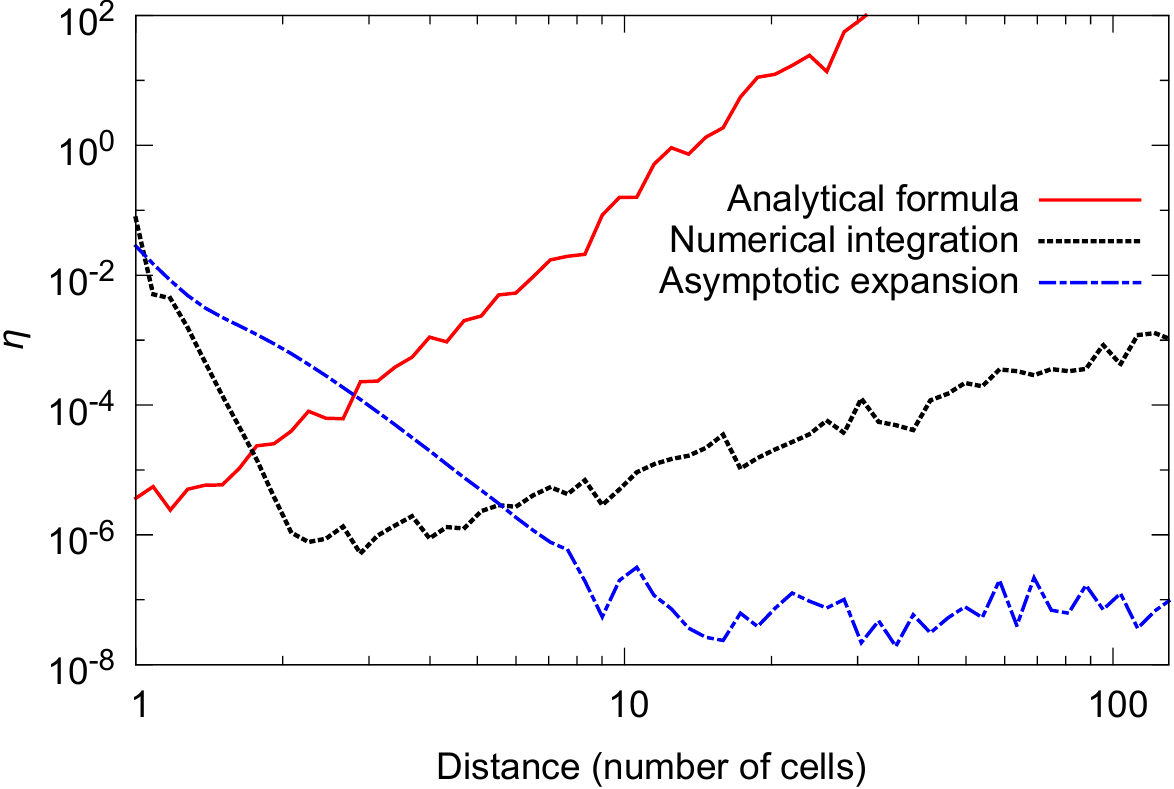}%
\caption{The relative error~$\eta$ of computing the demagnetizing tensor using single precision
arithmetic (precision $\sim 10^{-7}$), $1\!\times\! 1\!\times\! 1$ cell size, Kronrod-Patterson
sparse grid integration with full delay, order $k=7$.}%
\label{fig-single-precision} 
\end{figure}

The recent rise of General Purpose computing on Graphical
Processing Units (GPGPU) has re-invigorated single-precision floating
point operations: on these architectures single precision floating
point operations are generally much faster, and on cheaper cards the
only type of floating point operations provided. Additionally, the
available RAM on the GPU card is limited, providing another incentive
to use single rather than double precision floating point numbers.

We repeat the study presented in figure \ref{fig-final-comparison} but
use single precision numbers for all methods and show the results in figure
\ref{fig-single-precision}. The qualitative findings are the same as
for double precision numbers: the most accurate methods are as a
function of increasing distance: (i) the analytical formula, (ii) the
sparse grid numerical integration technique and (iii) the asymptotic
expression. 

However, the relevant cross-over points have moved to
shorter distances. The analytical expression becomes less accurate than 
numerical integration for more than 2 cells distance, and the
asymptotic expression is more accurate than numerical integration
for spacings greater than 8 cells.

As mentioned in \ref{sec:overview} we find that the analytical expression
for double precision (Figure \ref{fig-final-comparison}) provides only
4 significant digits (i.e. a relative error of $10^{-4}$) for a
distance of 100 cells. For single precision (Figure
\ref{fig-single-precision}) we find that the analytical expression
provides the same level of accuracy (i.e. 4 significant digits) only
for distances up to 3 cells. Correspondingly, the distance for which
the relative error exceeds 1, moves from over 300 cells with double
precision to 11 cells with single precision.

The accurate calculation of the demagnetizing tensor entries using
single precision floating point numbers only is challenging -- using the
best methods currently known and combining the three methods shown in
figure \ref{fig-single-precision}, the relative error can be kept
around or below $10^{-4}$.

For practical use of GPGPU single precision calculations for
micromagnetic simulation, we recommend to compute the demagnetizing
tensor using double-precision -- either on the GPU with some reduction
in speed if the GPU hardware supports this, or on the CPU with a more
significant time penalty. As the computation of the demagnetizing
tensor \eqref{eq-n} only needs to be done once and subsequent
computations of the demagnetizing field as required for energy
minimisation or time stepping only require to carry out the
convolution \eqref{eq:discreteconvolution}, it should be acceptable to
increase the accuracy of the demagnetizing tensor for a one-off time penalty in
the setup phase. 

\subsection{Forward and backward Fast Fourier Transform}
\label{sec:forw-backw-fast}

Our tests showed that the forward and inverse fast Fourier
transforms required to compute the convolution \eqref{eq:discreteconvolution} did
not introduce any significant numerical error in the calculation of
the demagnetizing field (either single or double precision).

\subsection{Other high accuracy methods}
\label{sec:other-high-accuracy}

We note for completeness that there are other options to compute the
demagnetizing tensor more accurately than any of the methods outlined
above if high accuracy is of utmost importance.

There are high precision arithmetic libraries available which provide software implementations of floating
point operations: the library user can choose the number of
significant digits used in the calculations (where 8 would
approximately correspond to single precision accuracy and 16 to double
precision floating point numbers). The larger the number of
significant digits to be used, the slower is the execution of these
operations. We have used this technique to obtain the reference data $\mathbf{N}_\text{exact}$
required to compute the error $\eta$ \eqref{eq:eta}. Using such libraries requires significant changes to source code, and
execution is extremely slow. It is impractical to use such
libraries for micromagnetic simulation.

\section{Summary}

We have compared the accuracy of computing the demagnetizing tensor
using the analytical formula, the asymptotic expansion, and numerical
integration. We obtain and provide quantitative data on the relative
error of demagnetizing tensor entries computed using all three methods.

We propose a new method using numerical integration to compute the
entries of the demagnetizing tensor which allows to increase the
accuracy from an error of $ 10^{-8}$ to an error of only $ 10^{-12}$ for
intermediate distances of between 4 and 80 simulation cells for the
commonly used double precision floating point numbers. In the context of
micromagnetic simulations, we find that the 7th order 641-point
Kronrod-Patterson sparse grid formula with full delay~\cite{Petras2003} and the
fixed-order 145-point rule~\cite{ecf,Sorevik85}  provide a reasonable tradeoff
between accuracy and performance. The integration points and weights for those
formulas are included in the supplementary information for this paper.

In the context of recent GPGPU use in micromagnetic simulations, we
also obtain accuracy data for the three methods using single precision
floating point numbers.

We thank Prof.~Ronald Cools for providing access to the Online Encyclopaedia of Cubature Formulas~\cite{ecf}. We acknowledge financial support from the EPSRC Centre for Doctoral Training grant EP/G03690X/1.

\bibliographystyle{elsarticle-num}
%\bibliography{numerical-integration}

\end{document}